\documentclass[aps,prd,superscriptaddress,preprint,nofootinbib]{revtex4-1}
\usepackage{amsmath}
\usepackage{graphicx}
\usepackage[active]{srcltx}
\def\beq{\begin{equation}}
\def\eeq{\end{equation}}

\def\beqn{\begin{eqnarray}}
\def\eeqn{\end{eqnarray}}

\begin{document}

\preprint{BRX-TH-655 }
\preprint{ Brown-HET-1634}

\title{    \hspace{-0.5cm} The Multi-Regge limit of NMHV Amplitudes in N=4 SYM Theory}

\author{Lev Lipatov}
\affiliation{St. Petersburg Nuclear Physics Institute, Russia}
\author{Alexander Prygarin}
\affiliation{Department of Physics, Brown University, Providence, RI 02912, USA}
\author{ Howard J. Schnitzer}
\affiliation{Theoretical Physics Group, Martin Fisher School of Physics
Brandeis University, Waltham, MA 02454, USA}

\begin{abstract}
We consider the multi-Regge limit for N=4  SYM  NMHV leading color amplitudes in two different formulations: the BFKL formalism for multi-Regge amplitudes in leading logarithm approximation, and superconformal N=4 SYM amplitudes. It is shown that the two approaches agree to two-loops for the $2 \to 4$  and  $3 \to 3$  six-point amplitudes. Predictions are made for the multi-Regge limit of three loop  $ 2 \to 4$ and   $3 \to 3$ NMHV amplitudes, as well as a particular sub-set of two loop $2 \to 2 + n$   N$^k$MHV amplitudes in the multi-Regge limit in the leading logarithm approximation from the BFKL  point of view.

\end{abstract}
\email{lipatov@mail.desy.de, prygarin@mail.desy.de, schnitzr@brandeis.edu}

 \maketitle

\tableofcontents

\section{Introduction}

The consideration of Regge behavior in Yang-Mills theories has a long history which began in the 1970's~\cite{Grisau}.  One particular application was motivated by a search for a description of the Pomeron  which respected unitarity. This led to the BFKL multi-Regge (MRK) formalism in leading logarithm approximation (LLA)~\cite{BFKL}. It turns out that the BFKL approach with adjoint exchange of Reggeized gluons~\cite{BLS2} is very well suited to the  discussion of  the remainder functions for  MHV amplitudes in the multi-Regge limit  in N=4 SYM theory, where the remainder function is defined as a contribution to be added to the BDS~\cite{BDS} amplitude. A natural extension of this issue is the analysis of the MRK limit for N$^k$MHV amplitudes, and NMHV amplitudes in particular. This will be a central theme of this paper.
More recently there have been enormous advances in techniques for computing leading color amplitudes in N=4 SYM theory~\cite{Dixon:2011xs, Brandhuber:2011ke, Bern:2011qt, Carrasco:2011hw, Ita:2011hi, Britto:2010xq, Schabinger:2011kb, Adamo:2011pv,  Elvang:2010xn, Drummond:2011ic, Henn:2011xk, Bargheer:2011mm, Bartels:2011nz}. An important step in this program was the BDS ansatz for all n-point functions  of planar  MHV amplitudes of N=4 SYM  theory~\cite{BDS}. The BDS ansatz was shown to be incomplete for $n > 5 $ point functions requiring a non-vanishing conformal invariant remainder function $R^{(n)}$ at two or more loops, as shown by explicit calculations~\cite{Goncharov:2010jf, Alday:2007he, BLS1,  BLS2, DelDuca:2009au,  DelDuca:2010zg,
Drummond:2007bm, Bern:2008ap, Drummond:2008aq}. Further the BDS amplitude for six or more point functions in the MRK limit does not have correct analyticity properties, as they do not exhibit certain Mandelstam cuts; those obtained from the BFKL equation for the $t$-channel exchange of two or more Reggeized gluons.  The MRK limit of the remainder function can be computed from the BFKL equation for MHV amplitudes in LLA in the MRK limit, which agrees with the explicit calculations in that limit.  This agreement encourages the application of the BFKL approach to N$^k$MHV  amplitudes, and comparison of the results to that of other methods when available.
In this paper we consider the MRK limit of NMHV amplitudes, as well as a particular subset of N$^k$MHV  amplitudes  obtained from the BFKL equations,  and compare these with results obtained by other methods. These agree whenever a comparison is possible, and leads to new predictions to be checked against future calculations. These BFKL results in the collinear, MRK limit exhibit close similarities with that of the OPE methods~\cite{Alday:2010ku, Gaiotto:2011dt, Sever:2011pc, Sever:2011da}, which therefore should be explored in more detail in the future.
 In this paper we review the BFKL kinematics and the remainder function for the $2 \to 4$ and $3 \to 3$ amplitudes in the MRK limit in Sec. 2.  A detailed analysis of the NMHV amplitude for the two-loop amplitude, and comparison to that of superconformal amplitudes is presented in Sec. 3. It is possible to include N$^k$MHV amplitudes for  more legs when two adjacent legs have their helicities flipped, which involve  simple modifications of the $2 \to 4$ case. Further extensions of these results will be be considered in forthcoming work.  Appendices present more details of the calculation.

\section{BFKL calculations }

\subsection{Kinematics for $2 \to 4$ }

\begin{figure}
\centering
\includegraphics[width=.5\textwidth]{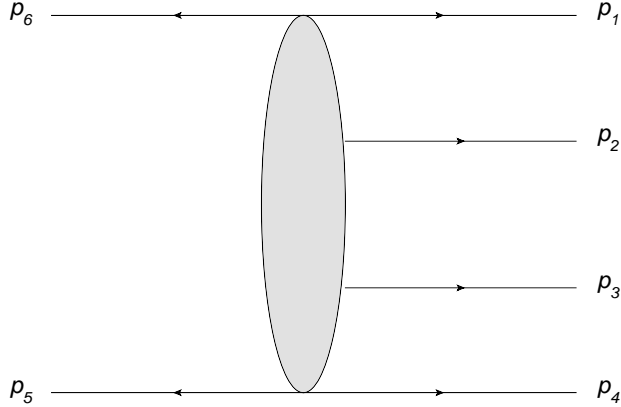}
\caption{ The $2 \to 4$ scattering amplitude in   Euclidean region.}
\label{fig:24}
\end{figure}

We consider multi-Regge kinematics~(MRK) of the $2 \to 4$ gluon  MHV amplitude for  $p_5+p_6 \to p_1 +p_2+p_3+p_4$  scattering depicted in Fig.~\ref{fig:24}.

In this kinematic  region   all $p^+$~( and $p^-$) components of the external particles are strongly ordered
\beqn
-p^+_6 \simeq p^+_1 \gg p^+_2  \gg p^+_3 \gg p^+_4 \simeq - p^+_5
\eeqn
with an inverse ordering for $p^-_{i}$.
Helicity  configurations throughout are with all momenta outgoing.
 Define the cross ratios
\beqn\label{cross}
u_{i}=\frac{x^2_{i,i+4} x^2_{i+1,i+3}}{x^2_{i,i+3} x^2_{i+1,i+4}}; \;\;\;\; i=1,2,3
\eeqn
with dual coordinates
\beqn
p_i=x_i -x_{i+1}.
\eeqn
The MRK limit becomes in the Euclidean region
\beqn\label{MRK24}
u_1 \to 1^{-}, \; u_2 \to 0^{+}, \; u_3 \to 0^{+}, \;\text{with} \;\; \tilde{u}_2 =\frac{u_2}{1-u_1} \simeq \mathcal{O}(1) \; \text{and}  \;
\tilde{u}_3 =\frac{u_3}{1-u_1} \simeq \mathcal{O}(1).
\eeqn
In general kinematics the remainder function has some square roots of the cross ratios in the arguments of the polylogarithms as it was shown   in  Ref.~\cite{Goncharov:2010jf}.
In the MRK limit  only two kinds of square roots survive,  and they can be  rationalized by choosing complex variables $w$ and $w^*$ (see Ref.~\cite{LP1}) related to the transverse momenta components of the produced particles
\beqn
w=\frac{(p_4+p_5) p_2}{(p_1+p_6)  p_3}, \;\;\;\; w^*=\frac{(p^*_4+p^*_5) p^*_2}{(p^*_1+p^*_6)  p^*_3}
\eeqn
or in terms of the cross ratios\footnote{The complex transverse momentum representation  is the primary  definition  of the complex variable $w$.   }
\beqn
w= \frac{1-\tilde{u}_2 -\tilde{u}_3 + \sqrt{(1-\tilde{u}_2 -\tilde{u}_3)^2 -4 \tilde{u}_2 \tilde{u}_3}}{2 \tilde{u}_2},
\;
w^*= \frac{1-\tilde{u}_2 -\tilde{u}_3 - \sqrt{(1-\tilde{u}_2 -\tilde{u}_3)^2 -4 \tilde{u}_2 \tilde{u}_3}}{2 \tilde{u}_2}. \nonumber
\eeqn

\subsection{The Remainder Function in the Mandelstam region}

\begin{figure}
\centering
\includegraphics[width=.5\textwidth]{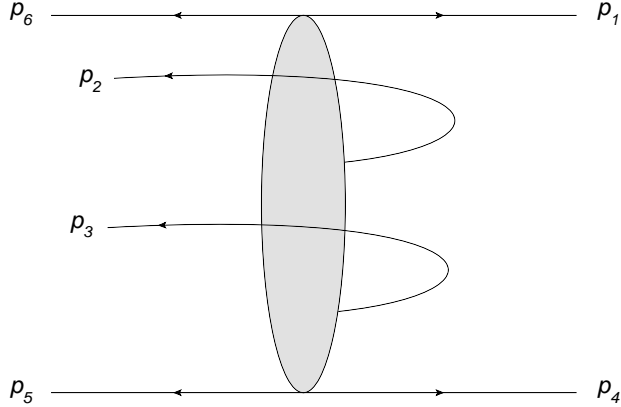}
\caption{ Mandelstam region for  $2 \to 4$   amplitude.}
\label{fig:24flip}
\end{figure}

The remainder function  is defined as a contribution to be added to the BDS amplitude.
In the Euclidean region it vanishes in   multi-Regge kinematics~\cite{BLS2, Brower:2008nm, Brower:2008ia}, but there are some regions, which we call Mandelstam regions where the remainder function have a divergent contribution of the order of $\log^{\ell-1} s $~($\ell$ is the number of loops). This happens due to the presence in those regions of so called Regge or Mandlestam cuts~\cite{BLS1, BLS2}, which are not accounted for by the BDS amplitude. The Mandlestam cuts are absent in the planar amplitudes  and manifest themselves only in non-planar cases.  The BDS ansatz is formulated for the planar amplitudes and we can make it non-planar in kinematics, while still being planar in color, flipping the produced particles as illustrated  in Fig.~\ref{fig:24flip}.
For the Mandelstam region   shown in Fig.~\ref{fig:24flip}  the remainder function was first calculated  to   leading logarithmic accuracy in Ref.~\cite{BLS2}
and given     to all orders by,
\beqn\label{Rmhv24}
R_{6;MHV}  \simeq \frac{i a  }{2} \sum_{n=-\infty}^{+\infty} \int_{-\infty}^{+\infty} d \nu
\frac{(-1)^n}{\nu^2 +\frac{n^2}{4}} w^{i\nu +\frac{n}{2}} (w^*)^{i\nu -\frac{n}{2}}
\left( \left(\frac{s_{45}}{s_0}\right)^{-a E_{n,\nu}}  -1 \right),
\eeqn
where $E_{n, \nu}$ is the eigenvalue of the BFKL equation in the adjoint representation
\beqn
E_{n, \nu} =-\frac{1}{2} \frac{|n|}{\nu^2 +\frac{n^2}{4}} +\psi\left(1+i \nu+\frac{|n|}{2}\right)
+\psi\left(1-i \nu+\frac{|n|}{2}\right)-2 \psi(1),
\eeqn
where $\psi (z)=\Gamma'(z)/\Gamma(z)$ and $a=g^2 N/8 \pi^2$.
The continuation to the Mandelstam region is $u_1= e^{-2\pi i} |u_1|$.
 The expression in eq.~(\ref{Rmhv24}) predicts the leading log remainder function to any loop order
\beqn
R_{6;MHV} = \sum_{\ell=2}^{\infty}a^{\ell} R^{(\ell)}_{6;MHV}.
\eeqn
At two loops it was calculated in Ref.~\cite{BLS2, Schabinger:2009bb}
\beqn\label{R26}
R^{(2)\; LLA}_{6;MHV}=- \frac{i  \pi  }{2}  \log   \left( \frac{s_{23}}{s_{0}} \right)
 \log   \left|1+w \right|^2  \log  \left|1+\frac{1}{w}\right|^2
\eeqn
and at three loops in Ref.~\cite{LP2}
\beqn\label{R36}
&&R^{(3)\; LLA}_{6;MHV}= \frac{i  \pi  }{4}  \log^2   \left( \frac{s_{23}}{s_{0}} \right)
\left(
 \log |w|^2 \log^2 |1+w|^2 -\frac{2}{3} \log^3 |1+w|^2  -\frac{1}{4} \log^2 |w|^2 \log |1+w|^2 \nonumber  \;\;\; \right.\\
 &&
 \left. \;\;\;\;
  +\frac{1}{2} \log |w|^2 \left( \text{Li}_2 \left(-w\right) +\text{Li}_2 \left(-w^*\right) \right)
 -\text{Li}_3 \left(-w\right)-\text{Li}_3 \left(-w^*\right)  \right)
\eeqn
At the leading logarithmic level the energy scale $s_0$ is arbitrary and among other possible choices we prefer
\beqn\label{scale24}
\frac{s_{23}}{s_0} =\frac{1}{\sqrt{u_2 u_3}},
\eeqn
which follows from the requirement of  Regge factorization and agreement with next-to-leading corrections at three loops as shown in Ref.~\cite{FadLip}.

In the MHV $2 \to 4 $ amplitude in   Regge kinematics,  the helicity of  the colliding particles is not changed, which limits the number of the possible helicity configurations to either $++++--$ or $+--+--$~(and their conjugates). In both cases the leading log remainder function is the same. For the NMHV case in MRK one can change a helicity of one of the produced particles $p_2$ or $p_3$,  and it is sufficient  to consider only one case, as all other cases are obtained  by complex conjugation of the complex $w$ variable.
Here we consider in more detail the  $++-+--$ helicity configuration, where the produced particle $p_3$ has  opposite helicity to that of the MHV case. The all orders  leading logarithm  NMHV remainder function  is  given by
\beqn\label{RNmhv24}
R^{LLA}_{6;NMHV}  \simeq   -\frac{i a }{2} \sum_{n=-\infty}^{+\infty} \int_{-\infty}^{+\infty} d \nu
\frac{(-1)^n}{(i\nu +\frac{n}{2})^2} w^{i\nu +\frac{n}{2}} (w^*)^{i\nu -\frac{n}{2}}
\left( \left(\frac{s_{23}}{s_0}\right)^{-a E_{n,\nu}}  -1 \right).
\eeqn
Note that eq.~(\ref{RNmhv24}) can be obtained from eq.~(\ref{Rmhv24}) by making the following substitution in the integrand
\beqn\label{subspm}
\frac{1}{-i\nu +\frac{n}{2}} \rightarrow -\frac{1}{i \nu +\frac{n}{2}},
\eeqn
which follows from the fact that to  leading order in MRK the impact factors for  gluons with opposite helicities are related by $\chi(\nu, n) \rightarrow \chi^* (-\nu, -n) $~(see Appendix~\ref{app:chi} for more details).
It also follows from the above property of the impact factors that to the leading logarithmic  order
\beqn \label{diffeq}
\int d w^* \frac{w}{ w^*} \frac{\partial }{\partial  w} R_{NMHV} = -R_{MHV}
\eeqn
for the helicity configuration under discussion.
The two loop NMHV remainder function in the leading logarithmic approximation is calculated in the Appendix~\ref{app:3loops} and given by
\beqn
&& R^{(2)\; LLA}_{NMHV} \simeq \frac{i \pi }{2} \log \left(\frac{s_{23}}{s_0}\right) \left\{ \frac{1}{1+w^*} \left( \frac{}{} \log |w|^2 \log (1+w^*)  -\text{Li}_2(-w)+\text{Li}_2(-w^*)
\right) \right.  \nonumber
\\
&&
\left.
+\frac{1}{1+\frac{1}{w^*}} \left(
 \log \frac{1}{|w|^2}  \log \left(1+\frac{1}{w^*}\right) -\text{Li}_2\left(-\frac{1}{w}\right)+\text{Li}_2\left(-\frac{1}{w^*}\right)\right)  \right\}
\eeqn

It is convenient to define the ratio function
\beqn\label{Pratio}
\mathcal{P}_{NMHV} =\frac{\mathcal{A}^{NMHV}}{\mathcal{A}^{MHV}},
\eeqn
which   to  leading logarithmic order can be written as
\beqn\label{PLLA}
\mathcal{P}^{LLA}_{NMHV}  \simeq R^{LLA}_{NMHV} -R^{LLA}_{MHV}
\eeqn
From  eq.~(\ref{Rmhv24}) and  eq.~(\ref{RNmhv24}) and the fact that
\beqn
-\frac{1}{(i \nu+\frac{n}{2})^2}-\frac{1}{\nu^2 +\frac{n^2}{4}}=-\frac{1}{\nu^2 +\frac{n^2}{4}} \frac{n}{ i \nu+\frac{n}{2} }
\eeqn
allows us to  write a compact expression for the ratio function in  leading logarithmic approximation
\beqn\label{PLLAint}
\mathcal{P}^{LLA}_{6\; NMHV}   \simeq  - \frac{i a }{2} \sum_{n=-\infty}^{+\infty} \int_{-\infty}^{+\infty} d \nu
\frac{(-1)^n}{\nu^2 +\frac{n^2}{4}}  \frac{n}{i\nu +\frac{n}{2}}w^{i\nu +\frac{n}{2}} (w^*)^{i\nu -\frac{n}{2}}
\left(\frac{s_{23}}{s_0}\right)^{-a E_{n,\nu}}.
\eeqn
Note that we dropped the minus unity in the brackets in eq.~(\ref{RNmhv24}) because, in  contrast to the remainder function,  the ratio function is well defined also at one loop. This is due to the fact that  the divergences at $n=\nu=0$ cancel  between the MHV and the NMHV parts resulting in a finite answer also at one loop
\beqn\label{P16}
\mathcal{P}_{6\;NMHV} ^{(1)\; LLA}&\simeq&  i \pi\frac{1  }{1+w^*} \log |1+w|^2 +  i \pi \frac{ w^*}{1+w^*} \log \left|1+\frac{1}{w}\right|^2
\\
&&  =-i\pi \frac{1}{1+w^*} \log \tilde{u}_2
-i\pi \frac{w^*}{1+w^*} \log \tilde{u}_3.  \;\;\; \nonumber
\eeqn

At two loops we have
\beqn\label{P26}
&& \mathcal{P}_{6\;NMHV} ^{(2)\; LLA} \simeq \frac{i \pi }{2} \log \left(\frac{s_{23}}{s_0}\right) \left\{ \frac{1}{1+w^*} \left( \frac{}{} \log |w|^2 \log (1+w^*)  -\text{Li}_2(-w)+\text{Li}_2(-w^*)
\right) \right.  \nonumber
\\
&&
\hspace{2cm}
\left.
+\frac{1}{1+\frac{1}{w^*}} \left(
 \log \frac{1}{|w|^2}  \log \left(1+\frac{1}{w^*}\right) -\text{Li}_2\left(-\frac{1}{w}\right)+\text{Li}_2\left(-\frac{1}{w^*}\right)\right)  \right\} \nonumber
 \\
 &&
 \hspace{2cm}
 +\frac{i \pi }{2} \log \left(\frac{s_{23}}{s_0}\right)\log   \left|1+w \right|^2  \log  \left|1+\frac{1}{w}\right|^2
\eeqn

\section{NMHV at two loops }

In this section  we consider the NMHV superamplitude   at two loops derived by Dixon, Drummond and Henn in Ref.~\cite{DixonNMHV}.
It is convenient to define the ratio function $\mathcal{P}$, which   relates all possible helicity configurations of the external particles to the MHV superamplitude
\beqn
\mathcal{A}=\mathcal{A}_{MHV} \times \mathcal{P}.
\eeqn

The expansion of $ \mathcal{P}$ in Grassmann  variables gives the corresponding type of amplitudes
\beqn
\mathcal{P}=1+\mathcal{P}_{NMHV}+\mathcal{P}_{N^2MHV}+...+ \mathcal{P}_{\overline{MHV}}.
\eeqn
We focus on the six particle amplitude, where at the tree level the ratio function can be expressed  in terms  of dual superconformal "R-invariants" as follows~\cite{Drummond:2008bq, Drummond:2008vq}
\beqn\label{treeNMHVr}
\mathcal{P}^{(0)}_{NMHV}=R_{1;35} +R_{1;36}+R_{1;46}.
\eeqn
It is useful to introduce the momentum twistors $Z_i$ and supertwistors $\mathcal{Z}_i$~\cite{Hodges:2009hk}
\beqn
\mathcal{Z}_i=(Z_i|\chi_i), \; Z_i^{R=\alpha, \dot{\alpha}}=(\lambda_i^{\alpha}, x_i^{\beta \dot{\alpha}} \lambda_{i\beta}), \; \chi_{i}^A=\theta_i^{\alpha A} \lambda_{i \alpha}
\eeqn
with
\beqn
(abcd)=  \epsilon_{RSTU} Z_{a}^R Z_{b}^S Z_{c}^T Z_{d}^U,
\eeqn
where one defines dual coordinates by
\beqn
p_i^{\alpha \dot{\alpha}}=\lambda_i^{\alpha}\tilde{\lambda}_i^{\dot{\alpha}}=x_{i}^{\alpha \dot{\alpha}} -x_{i+1}^{\alpha \dot{\alpha}}, \;\; q_i^{\alpha A}=\lambda_i^\alpha \eta_i^A=\theta_{i}^{\alpha A}-\theta_{i+1}^{\alpha A}.
\eeqn

The R-invariants can be  written compactly in terms of momentum twistors  using
\beqn\label{abcde}
[abcde]=\frac{\delta^4 ( \chi_a (bcde)+cyclic)}{(abcd)(bcde)(cdea)(deab)(eabc)}
\eeqn
and
\beqn
R_{r;ab}=[r,a-1, a, b-1, b].
\eeqn
For the six particle amplitude there are six different invariants.
For simplicity one can label them by $(t)$, using the momentum twistor $t$ that  is absent from the five arguments in the brackets, e.g.
\beqn
(1) \equiv [23456].
\eeqn
Using the identity between the invariants~\cite{Drummond:2008vq}
\beqn
(1)-(2)+(3)-(4)+(5)-(6)=0
\eeqn
we can write the NMHV amplitude (\ref{treeNMHVr}) as
\beqn\label{treeNMHVb}
\mathcal{P}^{(0)}_{NMHV}=(6)+(4)+(2)=(1)+(3)+(5).
\eeqn

The loop  contributions are taken into account   dressing   $(t)$ by functions of the dual conformal invariants $u_i$
\footnote{We use $u_i$ notation to avoid any confusion with complex variables $w$ and $w^*$. Our cross ratios are related to the ones in Ref.~\cite{DixonNMHV} by $u_1=u$, $u_2=v$ and $u_3=w$. The variables $y_i$ are identified as follows $y_1=y_u$, $y_2=y_v$ and $y_3=y_w$. }  \cite{DixonNMHV}
\beqn\label{Pgen}
\mathcal{P}_{NMHV} &&=\frac{1}{2} \left\{ \frac{}{}
\left[(1)+(4) \right]V(u_1, u_2,u_3)+\left[(2)+(5) \right]V( u_2,u_3, u_1)
+
\left[(3)+(6) \right]V(u_3, u_1, u_2) \right.  \nonumber
\\
&&
\left.
+\left[(1)-(4) \right]\tilde{V}(y_1, y_2,y_3)   -\left[(2)-(5) \right]\tilde{V} ( y_2,y_3, y_1)
+\left[(3)-(6) \right] \tilde{V}(y_3, y_1, y_2)
 \right\},
\eeqn
where
\beqn
y_i =\frac{u_i-z_{+}}{u_i-z_{-}}
\eeqn
are given  by
\beqn
z_{\pm}=\frac{1}{2} \left[-1+u_1+u_2+u_3 \pm \sqrt{\Delta}\right], \;\;\;
\Delta=(1-u_1-u_2-u_3)^2 -4 u_1 u_2 u_3.
\eeqn
The functions $V$ and $\tilde{V}$  represent parity conserving and parity violating amplitudes respectively and  obey the symmetry properties
\beqn\label{propVtV}
V(u_3, u_2, u_1)=V(u_1, u_2, u_3), \; \tilde{V}(y_3, y_2, y_1)=- \tilde{V}(y_1, y_2, y_3)
\eeqn
and are functions of the coupling constant
\beqn
V(a)=\sum_{\ell=0}^{\infty}a^\ell V^{(\ell)}, \;\; \tilde{V}(a)=\sum_{\ell=0}^{\infty}a^\ell \tilde{V}^{(\ell)}.
\eeqn

At tree level
\beqn
V^{(0)}=1, \; \tilde{V}^{(0)}=0
\eeqn
and at one loop we have~\cite{Schabinger:2011kb, Kosower:2010yk, Drummond:2008vq, Schabinger:2011wh}
\beqn
V^{(1)}(u_1, u_2,u_3)&& =\frac{1}{2} \left[ \frac{}{}
-\log u_1 \log u_3 +\log (u_1 u_3) \log u_2 +\text{Li}_2 (1-u_1)
\right.
\\
&&
\left.
\hspace{2cm}
+ \text{Li}_2 (1-u_2)
+\text{Li}_2 (1-u_3)-2 \zeta_2  \frac{}{}
\right],  \nonumber
\\
&& \hspace{-2.5cm}
\tilde{V}^{(1)}(u_i, u_j,u_k)=0,
\eeqn
while at two loops both  $V^{(2)}$ and $\tilde{V}^{(2)}$ are non-vanishing  and were calculated in Ref.~\cite{DixonNMHV}.
In the present study we check  the analytic properties of $V^{(1)}$, $V^{(2)}$ and $\tilde{V}^{(2)}$ going to the Mandelstam region and show that they correctly reproduce the BFKL calculations to   leading logarithmic accuracy.

\subsection{Multi-Regge kinematics in the  Euclidean region}

In this section we consider multi-Regge kinematics of eq.~(\ref{MRK24}) for $2 \to 4$ scattering and perform an analytic continuation of $V$ and $\tilde{V}$ to the corresponding Mandelstam region in Fig.~\ref{fig:24flip} reproducing the BFKL leading log calculations at    one and two loops.

In   multi-Regge kinematics $y_i$ are functions of only the complex variables $w$ and $w^*$
\beqn\label{yw}
y_1 \to 1, \; y_2 \to \tilde{y}_2 =\frac{1+w^*}{1+w}, \; y_3 \to \tilde{y}_3 =\frac{1+\frac{1}{w}}{1+\frac{1}{w^*}}.
\eeqn

Before discussing the Mandelstam region, we investigate the Regge behavior of the function $V$ and $\tilde{V}$ at one loop in the Euclidean region
\beqn
&& V^{(1)}(u_1, u_2, u_3) \simeq  \frac{1}{2} \log u_2 \log u_3, \\  \nonumber
&& V^{(1)}(u_3, u_1, u_2) \simeq   -\frac{1}{2} \log u_2 \log u_3, \\  \nonumber
&& V^{(1)}( u_2, u_3, u_1) \simeq   +\frac{1}{2} \log u_2 \log u_3   \nonumber
\eeqn
and two loops
\footnote{We calculate the asymptotics of $V$ and $\tilde{V}$ from the symbol in Ref.~\cite{DixonNMHV}, which captures only "pure" functions and not terms of lower transcendentality, such as $\zeta_2$ or those multiplied by a power of $\pi$.  }
\beqn
&& V^{(2)}(u_1, u_2, u_3) \simeq \frac{1}{16} \log^2 u_2 \log^2 u_3, \\  \nonumber
&& V^{(2)}(u_3, u_1, u_2) \simeq  -\frac{1}{16} \log^2 u_2 \log^2 u_3, \\ \nonumber
&& V^{(2)}( u_2, u_3, u_1) \simeq    \frac{1}{16} \log^2 u_2 \log^2 u_3  \\  \nonumber
&& \tilde{V}^{(2)}(y_i, y_j, y_k) \simeq 0.  \nonumber
\eeqn
We immediately notice a very disturbing feature of $V$, namely, that they are badly divergent in the MRK of eq.~(\ref{MRK24}) because
\beqn
\log u_2 \log u_3 \simeq \log^2 \delta + \log \delta \log (\tilde{u}_2 \tilde{u}_3)+\mathcal{O}(1),
\eeqn
where (see also eq.~(\ref{scale24}))
\beqn
 \delta=\sqrt{u_2 u_3 } =(1-u_1) \sqrt{\tilde{u_2}\tilde{ u_3 }} \to 0
\eeqn
is the only small parameter along with finite $\tilde{u}_2$ and $\tilde{u}_3$.
From   Regge theory we   expect all such divergences in the Euclidean region to cancel between them. This implies the condition
\beqn\label{cond}
[(1)+(4)]+[(2)+(5)]-[(3)+(6)]=0,
\eeqn
which we check   later by   explicit calculation of the R-invariant in eq.~(\ref{abcde}).
The function $\tilde{V}$ is zero at one loop and vanishing at two loops in MRK

\subsection{The Mandelstam Region}

We consider the Mandelstam region illustrated in Fig.~\ref{fig:24flip}, where we flip two produced particles. The analytic continuation for the $2 \to 4$ scattering, which takes one from the Euclidean region to the Mandelstam region  is given by
\beqn\label{analcont24}
u_1 =|u_1| e^{-i2 \pi}, \;u_2=|u_2|, \;  u_3=|u_3|.
\eeqn

After the analytic continuation eq.~(\ref{analcont24}) the functions $V$ at one loop in MRK read
\beqn
&& V^{(1)}(u_1, u_2, u_3)   \rightarrow V^{(1)}(u_1, u_2, u_3) +i\pi \log \delta-\frac{i 3 \pi}{2} \log \tilde{u}_2 +\frac{i \pi}{2} \log \tilde{u}_3
\\
&& V^{(1)}(u_3, u_1, u_2)   \rightarrow V^{(1)}(u_3, u_1, u_2) -i\pi \log \delta+\frac{i  \pi}{2} \log \tilde{u}_2 -\frac{i \pi}{2} \log \tilde{u}_3\nonumber \\
&&
V^{(1)}(u_2, u_3, u_1)   \rightarrow V^{(1)}( u_2, u_3, u_1) +i\pi \log \delta+\frac{i  \pi}{2} \log \tilde{u}_2 -\frac{i 3\pi}{2} \log \tilde{u}_3. \nonumber
\eeqn

We note that the cancelation of undesired $i\pi \log \delta$ terms is guaranteed by the condition eq.~(\ref{cond}) on the R-invariants. Indeed,  projecting $(t)$  onto the $++-+--$ helicity configuration, relevant for $2 \to 4 $ scattering in the multi-Regge kinematics, we obtain
\beqn\label{t24}
(1) \rightarrow \frac{1}{1+w^*} , \; (5) \rightarrow \frac{w^*}{1+w^*},   \;  (6) \rightarrow 1, \;   (2) \rightarrow 0, \;
(3) \rightarrow 0, \; (4) \rightarrow  0,
\eeqn
which agrees with the condition for the cancelation of large logarithms in eq.~(\ref{cond}).
Inserting  eq.~(\ref{t24}) in the ratio function $\mathcal{P}$ in eq.~(\ref{Pgen}) at tree level we get $\mathcal{P}=1$
\footnote{The tree level result was anticipated by Del Duca~\cite{DelDuca:1995zy}.}
 and at one  loop we reproduce the BFKL result of eq.~(\ref{P16}), namely
\beqn
\mathcal{P}_{6\;NMHV} ^{(1)\; LLA}&\simeq& -i\pi \frac{1}{1+w^*} \log \tilde{u}_2
-i\pi \frac{w^*}{1+w^*} \log \tilde{u}_3.  \;\;\;
\eeqn

At two loops the analytic continuation of $V$ and $\tilde{V}$ is performed using the prescription  for the symbol introduced in Ref.~\cite{Dixon3loops}.  The relative simplicity of the symbol after the analytic continuation allows us to find a corresponding function up to possible  "non-pure" functions, such as $\zeta_2$ or  those are built of powers of $\pi$ times pure functions, and thus are beyond the accuracy  of   the leading logarithmic approximation considered in the present study.  Each individual $V$ contains undesired large  logarithm terms of the order $i\pi \log^3 \delta$, but those cancel in sum leaving only  reasonable subleading logarithmic terms of the order $ i\pi \log^{\ell-1} \delta $~($\ell$ is the number of loops). The functions $\tilde{V}$ have only "good" leading log terms and thus we have
\beqn\label{Vtwo1}
   V^{(2)}(u_1, u_2, u_3) +V^{(2)}(u_3, u_1, u_2) && \rightarrow
 -\frac{i \pi}{2} \log \delta \log^2 \tilde{u}_2
 -\frac{i \pi}{2} \log \delta \log  \tilde{u}_2 \log  \tilde{u}_3+\mathcal{O}(1),
\eeqn
\beqn\label{Vtwo2}
   V^{(2)}( u_2, u_3,u_1) +V^{(2)}(u_3, u_1, u_2) && \rightarrow
 -\frac{i \pi}{2} \log \delta \log^2 \tilde{u}_3
 -\frac{i \pi}{2} \log \delta \log  \tilde{u}_2 \log  \tilde{u}_3  +\mathcal{O}(1), \nonumber
\eeqn
\beqn\label{Vtwo3}
\tilde{V}^{(2)}(y_1,y_2,y_3) \rightarrow  i \pi\log \delta  \left(  \frac{1}{2}  \log |w|^2 \log \frac{1+w}{1+w^*}
+ \text{Li}_2 (-w)-   \text{Li}_2 (-w^*)  \right) +\mathcal{O}(1). \nonumber
\eeqn

Inserting  eq.~(\ref{Vtwo1}) and eq.~(\ref{t24}) into the expression for the ratio function in eq.~(\ref{Pgen}) we get
\beqn\label{P26new}
 \mathcal{P}_{6\;NMHV} ^{(2)\; LLA} && = \frac{1}{2} \frac{1}{1+w^*} \left( \frac{}{}  V^{(2)}( u_1, u_2,u_3) +V^{(2)}(u_3, u_1, u_2)+\tilde{V}^{(2)}(y_1,y_2,y_3)-\tilde{V}^{(2)}(y_3,y_1,y_2)\right) \nonumber
\\
&& \hspace{-1cm}
+
\frac{1}{2} \frac{w^*}{1+w^*} \left( \frac{}{}  V^{(2)}( u_2, u_3,u_1) +V^{(2)}(u_3, u_1, u_2)+\tilde{V}^{(2)}(y_2,y_3,y_1)-\tilde{V}^{(2)}(y_3,y_1,y_2)\right)  \;\;\;\; \nonumber
\\
&& \hspace{-1cm} \simeq - \frac{i \pi }{2} \log  \delta \left\{ \frac{1}{1+w^*} \left( \frac{}{}- \log |w|^2 \log (1+w)+\log^2 |1+w|^2  -\text{Li}_2(-w)+\text{Li}_2(-w^*)
\right) \right.  \nonumber
\\
&& \hspace{-1cm}
\left.
+\frac{1}{1+\frac{1}{w^*}} \left(
- \log \frac{1}{|w|^2}  \log \left(1+\frac{1}{w }\right) +\log^2 \left|1+\frac{1}{w}\right|^2 -\text{Li}_2\left(-\frac{1}{w}\right)+\text{Li}_2\left(-\frac{1}{w^*}\right)\right)  \right\}
\eeqn
which reproduces the BFKL result  in eq.~(\ref{P26}). The overall minus sign is due to the fact that $\log (s_{23}/s_0)=- \log \delta$.

It is worth emphasizing that the first two lines of eq.~(\ref{P26new}) represent a general structure valid to any loop order in the leading logarithm approximation. It is unambiguously fixed by properties of $V$ and $\tilde{V}$ in eq.~(\ref{propVtV}), the target-projectile symmetry as well as the proper collinear limit as follows. In the multi-Regge kinematics of eq.~(\ref{MRK24}) the variables $u_i$ and $y_{i}$ can be written as
\beqn\label{MRKwuy}
u_{1} \to 1, \; u_{2} \to \frac{1-u_1}{|1+w|^2}, \; u_{3} \to \frac{1-u_1}{|1+\frac{1}{w}|^2},\; y_1 \to 1, \; y_2 \to \frac{1+w^*}{1+w}, \; y_3 \to \frac{1+\frac{1}{w}}{1+\frac{1}{w^*}}.
\eeqn
 The target-projectile symmetry means  that the result is invariant under an exchange  of the colliding particles in Fig.~\ref{fig:24flip}, which implies\footnote{Here we took into account the fact that the produced particles having $+-$ helicity configuration should have $-+$ helicities to preserve the target-projectile symmetry after the exchange of the colliding particles.  }  $w \leftrightarrow 1/w^*$    and thus
  \beqn
 u_1 \leftrightarrow u_1, \; u_2 \leftrightarrow u_3, \; y_2 \leftrightarrow y_3.
 \eeqn
This fixes the combination of $V$ and $\tilde{V}$ in the brackets  in eq.~(\ref{P26new}), but leaves some freedom of assigning this combination to either $1/(1+w^*)$ or $1/(1+\frac{1}{w^*})$. It is resolved by demanding of a proper collinear limit, i.e. any function which multiplies $1/(1+w^*)$ should vanish for $|w| \to 0$. The overall coefficient is fixed by the tree level expression. The above arguments are valid to any loop order  determining the general structure of the first two lines in eq.~(\ref{P26new}) for $++-+--$ helicity configuration.

In a similar way we checked the other $+-++--$ helicity configuration for the $2 \to 4$ amplitude as well, and found the analytic continuation of $V$ and $\tilde{V}$  to be consistent with BFKL calculations for the $ 3 \to 3$ amplitude for helicity configurations $+++---$, $+-+-+-$ and their conjugates. In the leading logarithm approximation the $3 \to 3$ case differs from the $2 \to 4$ case only by the   sign
 \beqn
 \mathcal{P}^{LLA}_{2 \to 4\; NMHV}=-\mathcal{P}^{LLA}_{3 \to 3\; NMHV}=\mathcal{P}^{LLA}_{6\; NMHV}
 \eeqn
  for $\mathcal{P}^{LLA}_{6\; NMHV}$  in eq.~(\ref{PLLAint}). The same is true also for the MHV and NMHV remainder functions.

 The analytic continuation  to the Mandelstam region for the $3 \to 3 $ amplitude  is given by~\cite{BLP3to3, BLS2}
\beqn
u_1 =|u_1| e^{ i2\pi }, \;u_2 =|u_2| e^{ -i\pi }, \; u_3 =|u_3| e^{ -i\pi }
\eeqn
and the multi-Regge kinematics reads
\beqn\label{MRK33}
u_1 \to 1^{+}, \; u_2 \to 0^{+}, \; u_3 \to 0^{+}, \;\text{with} \;\; \tilde{u}_2 =\left|\frac{u_2}{1-u_1}\right| \simeq \mathcal{O}(1) \; \text{and}  \;
\tilde{u}_3 =\left|\frac{u_3}{1-u_1} \right| \simeq \mathcal{O}(1).
\eeqn
 Note that in the $3 \to 3$ case $1-u_1$ is negative resulting in the difference of the real part between $3 \to 3$ and $2 \to 4 $ remainder functions as discussed in the next section.

Using the property in eq.~(\ref{diffeq}) and the three loop remainder function in eq.~(\ref{R36}) we calculate  the three loop leading log remainder  function in eq.~(\ref{RNmhv24})
\beqn\label{R3NMHVLLA}
R^{(3)\; LLA}_{6;NMHV}  && \simeq \frac{i \pi }{4} \log^2 \delta \left(
\frac{1}{1+w^*} f_3(w,w^*)
+
\frac{1}{1+\frac{1}{w^*}} f_3\left(\frac{1}{w},\frac{1}{w^*}\right)
\right),
\eeqn
where
\beqn\label{f3}
f_3(w,w^*)=&& -\frac{1}{2} \log^2 |w|^2 \log(1+w^*)+\log (-w) \left( \log^2 (1+w^*)-\log^2 (1+w) \frac{}{}\right) \\
&& +2 \zeta_2 \log |1+w|^2 +\frac{1}{2} \log |w|^2 \left( \text{Li}_2 (-w)-\text{Li}_2 (-w^*)\frac{}{}\right) \nonumber
\\
&&
-2 \log |1+w|^2 \text{Li}_2 (-w) -2 \text{Li}_3 (1+w)-2 \text{Li}_3 (1+w^*)+ 4 \zeta_3.  \nonumber
\eeqn
The leading log ratio function is then found from  eq.~(\ref{R36}) and eq.~(\ref{PLLA})
\beqn\label{P3NMHVLLA}
 \mathcal{P}^{(3)}_{6\;NMHV} \simeq R^{(3)\; LLA}_{6;NMHV} -R^{(3)\; LLA}_{6;MHV}
\eeqn

There is an ambiguity related to the indefinite integral in eq.~(\ref{diffeq}), which is resolved by demanding   the singlevaluedness  and   proper collinear behavior of $R^{(3)\; LLA}_{NMHV}$ (see  Appendix~\ref{app:3loops} for more details).

Eqs.~(\ref{R3NMHVLLA})-(\ref{P3NMHVLLA}) present a prediction to be checked against an eventual 3-loop calculation of $V^{(3)}$ and $\tilde{V}^{(3)}$.

\section{Real part of the remainder function}
In this section we calculate the real part of the remainder function at the next-to-leading logarithm order.
The leading logarithm contribution to the remainder function is pure imaginary, of the order $ \log^{\ell-1} \delta$~($\ell$ is the number of loops)
and comes entirely from the Mandelstam cut. The real part appears only at the next-to-leading logarithm level of the order of $\log^{\ell-2}$ and originates from both Mandelstam cuts and Regge poles as it was shown in Ref.~\cite{LipDisp}. There is no full separation between poles and cuts in the remainder function due to the fact that the BDS amplitude, lacking the entire contributions from Mandelstam cuts, still has some residual terms which can be assigned to Mandelstam cuts. Those are removed from the remainder function by introducing a phase $\Delta$ extracted from the BDS amplitude at one loop
\beqn
\Delta =\frac{\gamma_K}{8} \log \tilde{u}_2 \tilde{u}_3 =\frac{\gamma_K}{8} \log \frac{|w|^2 }{|1+w|^4},
\eeqn
where $\gamma_K \simeq 4 a $ and $ a = g^2 N_c/8 \pi^2 $ are the cusp anomalous dimension and the coupling constant respectively.
Thus one can write~\cite{LipDisp, LP2} the dispersion relation for the real and imaginary part of the remainder function for the $2 \to 4 $ amplitude
\beqn\label{disp24}
e^{i \pi \Delta}   R_{2 \to 4} =\cos  \pi \omega_{ab} + i \int_{- i\infty}^{+i\infty} \frac{d \omega }{2 \pi i } f (\omega) e^{-i \pi \omega } |\delta|^{-\omega}
\eeqn
and the $3 \to 3 $ scattering
\beqn\label{disp33}
 e^{-i \pi \Delta}  R_{3 \to 3} =\cos  \pi \omega_{ab} - i \int_{- i\infty}^{+i\infty} \frac{d \omega }{2 \pi i } f (\omega)  |\delta|^{-\omega}.
\eeqn
The phase $\Delta$ removes the residual cut terms  of the BDS amplitude from the remainder function and the last term in  eq.~(\ref{disp24}) and  eq.~(\ref{disp33}) restores the correct Mandelstam cut contribution. The Regge poles are accounted for by the $\cos  \pi \omega_{ab}$ term with \beqn
\omega_{ab}= \frac{\gamma_K}{8}  \log \frac{\tilde{u}_3}{   \tilde{u}_2}= \frac{\gamma_K}{8}  \log  |w|^2.
\eeqn

To the leading logarithm order  for the MHV amplitude the function $f(\omega)$ is determined from  eq.~(\ref{Rmhv24}) as
 \beqn\label{Rmhv24app}
  i \int_{- i\infty}^{+i\infty} \frac{d \omega }{2 \pi i } f^{LLA}_{MHV} (\omega) |\delta|^{-\omega}=\frac{i a  }{2} \sum_{n=-\infty}^{+\infty} \int_{-\infty}^{+\infty} d \nu
\frac{(-1)^n}{\nu^2 +\frac{n^2}{4}} w^{i\nu +\frac{n}{2}} (w^*)^{i\nu -\frac{n}{2}}
\delta^{a E_{n,\nu}},
\eeqn
 and the next-to-leading corrections to $f(\omega)$ were found in Refs.~\cite{LP2, FadLip}.
The phase coefficient $e^{- i\pi \omega}$ in the integrand of eq.~(\ref{disp24}) makes the real part of remainder function to obtain contributions from the imaginary part substituting
\beqn
 \log \delta  \to   \log \delta+i \pi
\eeqn
in the leading logarithm terms of the same loop order, whereas the phase $\Delta$  gives  the contribution to the real part from the imaginary leading logarithm terms of the previous loop order. For example, expanding  eq.~(\ref{disp24}) to the second order in $a$ one obtains
\beqn\label{2loopsdisp24}
a^2 \Re \left(R^{(2); NLLA}_{2 \to 4; \; MHV }\right) \simeq   -i \pi \Delta  R^{(1); LLA}_{2 \to 4; \; MHV } -\frac{\pi^2 \omega^2_{ab}}{2}+\frac{\pi^2 \delta^2 }{2}+R^{(2);  LLA}_{2 \to 4; \; MHV }|_{\log \delta \to i \pi} \simeq 0
\eeqn
provided the one loop remainder function $R^{(1); LLA}_{2 \to 4; \; MHV }$ is set to be zero. In eq.~(\ref{2loopsdisp24}) the Regge pole and Mandelstam cut contributions cancel out  resulting in the  zero real part for the MHV remainder at two loops for the $2 \to 4$ scattering amplitude. For the $3 \to 3 $ case  in eq.~(\ref{disp33}) the mixing phase $e^{-i \pi \omega }$ is absent in the integrand,  and thus we have a non-vanishing real part
\beqn\label{2loopsdisp33}
a^2 \Re \left(R^{(2); NLLA}_{3 \to 3; \; MHV }\right) \simeq   -i \pi \Delta  R^{(1); LLA}_{3 \to 3; \; MHV } -\frac{\pi^2 \omega^2_{ab}}{2}+\frac{\pi^2 \delta^2 }{2}= \frac{\pi^2}{2} \log |1+w|^2 \log \left|1+\frac{1}{w} \right|^2. \;\;\;
\eeqn
The absence of $e^{-i \pi \omega }$ in eq.~(\ref{disp33}) allows us to make all loop prediction for the following object~\cite{BLP3to3}
\beqn
\Re \left(e^{ -i \pi \Delta } R_{ 3 \to 3}\right) =\cos \pi \omega_{ab},
\eeqn
which is valid also in the strong coupling region.

The dispersion relations eq.~(\ref{disp24}) and eq.~(\ref{disp33}) remain valid also for the NMHV remainder functions, because in the derivation no assumption was made about the helicity configuration of the produced particles. The real part of the $2 \to 4$ next-to-leading ratio function is given by
\beqn\label{ReP24}
\Re ( \mathcal{P}_{2 \to 4}^{NLLA}) && \simeq -i\pi \Delta \mathcal{P}_{2 \to 4}^{LLA}+\mathcal{P}_{2 \to 4}^{LLA}|_{\log \delta \to \log \delta +i \pi}-\mathcal{P}_{2 \to 4}^{LLA}
\\
&&
\simeq -i\pi \Delta \mathcal{P}_{6}^{LLA}+\mathcal{P}_{6}^{LLA}|_{\log \delta \to \log \delta +i \pi}-\mathcal{P}_{6}^{LLA}, \;\;\; \nonumber
\eeqn
and for $3 \to 3$ scattering it reads
\beqn\label{ReP33}
\Re ( \mathcal{P}_{3 \to 3}^{NLLA}) \simeq i\pi \Delta \mathcal{P}_{3\to 3}^{LLA}\simeq -i\pi \Delta \mathcal{P}_{6}^{LLA}
\eeqn
for the leading logarithm ratio function $\mathcal{P}_{6}^{LLA}$ in eq.~(\ref{PLLAint}).
The contributions of the Regge poles $\cos \omega_{ab}$ completely cancel out in the ratio function having the same sign in  eq.~(\ref{2loopsdisp24}) and eq.~(\ref{2loopsdisp33}).

The last two terms in  eq.~(\ref{ReP24}) can be written as
\beqn
\hspace{-1cm}\mathcal{P}_{6}^{LLA}|_{\log \delta \to \log \delta +i \pi}-\mathcal{P}_{6}^{LLA}
 \simeq  \frac{ a^2 \pi }{2} \sum_{n=-\infty}^{+\infty} \int_{-\infty}^{+\infty} d \nu
\frac{(-1)^n}{\nu^2 +\frac{n^2}{4}}  \frac{n}{i\nu +\frac{n}{2}}w^{i\nu +\frac{n}{2}} (w^*)^{i\nu -\frac{n}{2}}
E_{n,\nu} \;\delta^{a  E_{n,\nu}}.
\eeqn
We checked that the $V$ and $\tilde{V}$ of Ref.~\cite{DixonNMHV} correctly reproduce the real parts of the $2 \to 4$ and $3 \to 3$ remainder functions at two loops.
Expressions in eq.~(\ref{ReP24}) and eq.~(\ref{ReP33}) together with eqs.~(\ref{P26new})-(\ref{P3NMHVLLA})  give a prediction for the next-to-leading real part of the remainder at  three loops.

\section{More legs}
In this section we consider N$^k$MHV in the leading log approximation with more external gluons.
We start our discussion with the $2 \to 5$ amplitude, where we have three produced particles with momenta $p_2$, $p_3$ and $p_4$.  In  multi-Regge kinematics the number of possible helicity configurations is limited due to the fact that the colliding particles have eikonal vertices and as a result their helicities stay the same. In the convention where all momenta are outgoing  this implies that helicities of particles with $p_1$ and $p_7$ should have opposite sign, and the same for gluons with $p_5$ and $p_6$. The helicities of the produced particles with momenta $p_4$, $p_5$ and $p_6$ are arbitrary. The $2 \to 5$ MHV amplitude was considered in Ref.~\cite{Bartels25, Prygarin:2011gd} and it was shown that its leading log  remainder function can be written as a sum of two $2 \to 4$ remainder functions. This happens due to some cancelations between propagators and effective vertices for particles of the same helicity. Unfortunately this is not the case with NMHV amplitudes, but one can consider Mandelstam  regions, where only two adjacent particles are flipped and then the remainder function is given by the same expression as for $2 \to 4$ case in eq.~(\ref{RNmhv24}) with redefined $ s_{23}/s_0 $ and $w$.
\begin{figure}
\centering
\includegraphics[width=.5\textwidth]{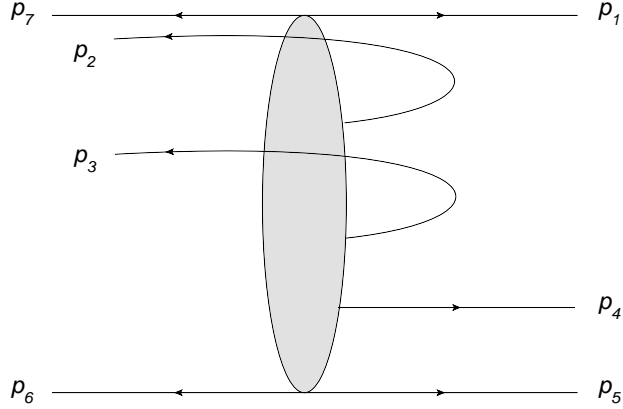}
\caption{ One of the  Mandelstam regions for $2 \to 5$ amplitude in which the remainder function reduces to the $2 \to 4$ case with redefined momenta.}
\label{fig:25flip}
\end{figure}
For example, when we flip produced particles with momenta $p_2$ and $p_3$ as depicted in Fig.~\ref{fig:25flip} for $++-++--$ helicity configuration we get
\beqn\label{RNmhv24xx}
R^{LLA;+-}_{7;2;NMHV}  \simeq   -\frac{i a }{2} \sum_{n=-\infty}^{+\infty} \int_{-\infty}^{+\infty} d \nu
\frac{(-1)^n}{(i\nu +\frac{n}{2})^2} w'^{i\nu +\frac{n}{2}} (w'^*)^{i\nu -\frac{n}{2}}
\left( \left(\frac{s_{45}}{s'_0}\right)^{-a E_{n,\nu}}  -1 \right),
\eeqn
where the cross ratios are as in eq.~(\ref{cross}), but with $i=1$ to $7$;
\beqn
\frac{s_{23}}{s'_0}=\frac{1}{\sqrt{ u_{6}  u_{7}}},
\eeqn
and\footnote{Here $p_i$ denotes the complex transverse momenta.}
\beqn
w'\equiv \frac{(p_5+p_6) p_2}{(p_7+p_1)  p_3}=\frac{1-\tilde{u}_{6} -\tilde{u}_{7} + \sqrt{(1-\tilde{u}_{6} -\tilde{u}_{7})^2 -4 \tilde{u}_{6} \tilde{u}_{7}}}{2 \tilde{u}_{6}}
\eeqn
for
\beqn
\tilde{u}_{1}=\frac{u_{6}}{1-u_{1}}, \;\; \tilde{u}_{7}=\frac{u_{7}}{1-u_{1}}.
\eeqn
The multi-Regge kinematics for $2 \to 5$ scattering implies
\beqn
1-u_1 \propto \delta, \; 1-u_2 \propto \delta,  \;1-u_5 \propto \delta^2, \; \delta \to 0
\eeqn
and the rest of cross ratios are of the order of $\delta$.

In this case the ratio function to   leading log accuracy  is   given by
\beqn
\mathcal{P}^{LLA;+-}_{7;2;NMHV}   \simeq  - \frac{i a }{2} \sum_{n=-\infty}^{+\infty} \int_{-\infty}^{+\infty} d \nu
\frac{(-1)^n}{\nu^2 +\frac{n^2}{4}}  \frac{n}{i\nu +\frac{n}{2}}w'^{i\nu +\frac{n}{2}} (w'^*)^{i\nu -\frac{n}{2}}
\left(\frac{s_{23}}{s'_0}\right)^{-a E_{n,\nu}}.
\eeqn
The corresponding analytic continuation reads
\beqn
u_{1}=|u_{1}| e^{-i2\pi}, \; u_{3}=|u_{3}| e^{-i\pi}, \; u_{4}=|u_{4}| e^{+i\pi},
\eeqn
while the other   cross ratios remain  the same.

In a similar way we can find  the  ratio function for many other Mandelstam regions, where  only two adjacent particles are flipped. The problem reduces to a proper redefinition of the energy scale $s_0$, the complex variable $w$ and the analytic continuation done case by case.

The NMHV superamplitudes for $n=7$ were considered in Refs.~\cite{Bern:2004ky, Drummond:2008bq, Korchemsky:2009hm}, which can be analyzed analogously to that of $n=6$. This is a project for future work.
It worth emphasizing that the ratio function for the Mandelstam regions, where we flip two adjacent particles, in the leading order does not depend on the helicities of the all other particles. More detailed discussion on this topic will be presented by us elsewhere.

\section{Conclusions}

  The multi-Regge limit (MRK)  for N=4 SYM NMHV amplitudes were considered in two different
formulations:
the BFKL formalism for multi-Regge amplitudes in leading logarithm approximation, and
superconformal N=4 SYM amplitudes.
It was shown that the  two approaches agreed in explicit calculations in leading logarithm approximation to two loops for the six-point gluon amplitudes. Predictions
were made for three –loop six point NMHV amplitudes and two-loop seven-point
NMHV amplitudes in leading logarithm approximation from the BFKL point of view.
Comparisons with similar calculations from superconformal amplitudes should strengthen the connection between these two methods.
  Another approach to computing the remainder functions is that of the operator product expansion (OPE) developed by Alday, Gaiotto, Maldacena, Sever, and Vieira
(AGMSV)~\cite{Alday:2010ku, Gaiotto:2011dt, Sever:2011pc, Gaiotto:2010fk}. In particular Sever, Vieira, and Wang~\cite{ Sever:2011da} rederive the one-loop NMHV six-point amplitudes from the OPE point of view. There appears to be a connection between the OPE methods and the BFKL results when compared in the collinear,
multi-Regge limits as shown in Ref.~\cite{Bartels:2011xy}. It would be interesting to find the precise relationship between the two points of view, as this could offer additional insights into this class of problems.

 After this paper was posted  Dixon, Duhr and Pennington~\cite{Dixon:2012yy} presented an extension of the $2 \rightarrow  4$ MHV and NMHV amplitudes in the MRK limit
to 10-loops using single-valued harmonic polylogarithms as a basis. They confirm the results of this paper.

\section{Acknowledgements}
We thank J.~Bartels, S.~Caron-Huot, L.~Dixon, V.~S.~Fadin, J.~Henn, G.~P.~Korchemsky, A.~Kormilitzin,  E.~M.~Levin,  A.~Sever, M.~Spradlin, C.~-I~Tan, C.~Vergu and A.~Volovich for
helpful discussions. The work of AP is supported in part by the the US National Science Foundation under
Grant PHY-064310 PECASE.
The research of HJS is supported in part by the Department of Energy under Grant DE-FG02-92ER40706.

\newpage

\appendix

\setcounter{equation}{0}

\renewcommand{\theequation}{A.\arabic{equation}}

\section{Impact Factors in the BFKL Approach}\label{app:chi}

In this section we find  the leading order  impact factor needed for  calculating the NMHV amplitude in leading logarithm approximation.  We adopt the momenta convention  of Ref.~\cite{BLS2} because our analysis is tightly related to   discussion presented   in Chapters 2 and 5 of Ref.~\cite{BLS2}.
Firstly we note that the effective production vertex $C_{\mu }(\mathbf{q}_2, \mathbf{q}_1)$ can be written in a very compact way for a definite helicity of the produced particles
\beqn
C_{\mu }(\mathbf{q}_2, \mathbf{q}_1) e_{\mu} (\mathbf{k}_1) =\sqrt{2} \frac{q_2 q^*_1}{k_1},
\eeqn
where we introduce the complex transverse momenta
\beqn
k=k_x+i k_y, \; \; k^* =k_x -i k_y.
\eeqn
Following the lines of Chapter 2 of Ref.~\cite{BLS2} we readily find that the impact factors for the opposite helicities are related by complex conjugation in   momentum space~(see eq.~(13) of Ref.~\cite{BLS2}) and thus we have
\beqn
\Phi^+_{2}= \frac{k_2 (k^{'}-q_2)}{q_2 (k^{'}-k_2)}, \;\;
\Phi^-_{2}= \frac{k^*_2 (k^{*'}-q^*_2)}{q^*_2 (k^{*'}-k^*_2)}.
\eeqn
The impact factor has to be convolved with the BFKL Green function (see eqs.~(84)-(92) of Ref.~\cite{BLS2})
\beqn\label{convapp}
\chi^{\pm}_2 = \int \frac{d^2 k' }{2 \pi } \frac{|q_2|^2 }{|k'|^2 |q_2 -k'|^2 }
\left(  \frac{q_2 -k'}{k'}\right)^{i \nu +\frac{n}{2}}
\left(  \frac{q^*_2 -k^{*'}}{k^{*'}}\right)^{i \nu -\frac{n}{2}} \Phi^\pm_{2}.
\eeqn
It is easy to see from eq.~(\ref{convapp}) that the impact factors for different helicities are related by
\beqn\label{conjapp}
\chi^{-} (\nu, n)= (\chi^{+}(-\nu, -n) )^*
\eeqn
rather than a simple conjugation.
The $\chi_2^{+}$ in eq.~(\ref{convapp}) was calculated in Ref.~\cite{BLS2}
\beqn
\chi_2^{+}=-\frac{1}{2} \frac{1}{i \nu -\frac{n}{2}}
\left(\frac{q_3^*}{k_2^*}\right)^{i \nu -\frac{n}{2}}
\left(\frac{q_3}{k_2}\right)^{i \nu +\frac{n}{2}}
\eeqn
and eq.~(\ref{conjapp}) implies eq.~(\ref{subspm}) resulting in  difference  of  the integral representations of the leading logarithm MHV and NMHV remainder functions in eq.~(\ref{Rmhv24}) and eq.~(\ref{RNmhv24}) respectively.

\setcounter{equation}{0}

\renewcommand{\theequation}{B.\arabic{equation}}
\section{Leading Logarithm  NMHV Remainder  Functions at One, Two and Three  Loops}\label{app:3loops}

In this section we calculate the leading logarithm ratio function in eq.~(\ref{PLLAint}) given by
\beqn\label{PLLAapp}
\mathcal{P}^{LLA}_{6\; NMHV}   \simeq  - \frac{i a }{2} \sum_{n=-\infty}^{+\infty} \int_{-\infty}^{+\infty} d \nu
\frac{(-1)^n}{\nu^2 +\frac{n^2}{4}}  \frac{n}{i\nu +\frac{n}{2}}w^{i\nu +\frac{n}{2}} (w^*)^{i\nu -\frac{n}{2}}
\left(\frac{s_{23}}{s_0}\right)^{-a E_{n,\nu}}.
\eeqn
In contrast to the remainder function for the NMHV amplitude in eq.~(\ref{RNmhv24}), the ratio function is finite even at one loop because the IR divergences cancel between the MHV and NMHV remainder functions. Technically, the divergence of the type $\int d \nu / \nu^2$ is absent here due to the presence of $n$ in the numerator, which makes the whole expression to vanish at $n=0$.

We start with the one loop case
\beqn\label{PLLAapp1}
\mathcal{P}^{(1)\;LLA}_{6\; NMHV}   \simeq  - \frac{i  }{2} \sum_{n=-\infty}^{+\infty} \int_{-\infty}^{+\infty} d \nu
\frac{(-1)^n}{\nu^2 +\frac{n^2}{4}}  \frac{n}{i\nu +\frac{n}{2}}w^{i\nu +\frac{n}{2}} (w^*)^{i\nu -\frac{n}{2}}
\eeqn
 and calculate $\mathcal{P}^{(1)\;LLA}_{6\; NMHV}$ using the Cauchy theorem as follows. We assume $|w| >1$ and close the integration contour in the upper semiplane. Then we have poles at $ \nu =i n/2$ for which $n>0$,  and  poles at
$ \nu =-i n/2$ for which $n<0$.
The residues at  poles $ \nu =i n/2$ give
\beqn
-i\pi \sum_{n=1}^{\infty}\frac{(w^*)^{-n}}{n} (1+ n\log |w|^2)=  i\pi  \log(1+w^*)
+\frac{i\pi}{1+w^*} \log w -\frac{i\pi \; w^* }{1+w^*} \log w^* ,
\eeqn
while from poles at $ \nu =-i n/2$ we have
\beqn
i\pi \sum_{n=-\infty}^{-1}\frac{w^{n}}{n} = i\pi  \log\left(1+\frac{1}{w}\right).
\eeqn
Adding the two contributions we have
\beqn\label{P16app1}
\mathcal{P}_{6\;NMHV} ^{(1)\; LLA}&\simeq&  i \pi\frac{1  }{1+w^*} \log |1+w|^2 +  i \pi \frac{ w^*}{1+w^*} \log \left|1+\frac{1}{w}\right|^2.
\eeqn

We notice that $\mathcal{P}_{6\;NMHV} ^{(1)\; LLA}$ can be written as
\beqn\label{P16app2}
\mathcal{P}_{6\;NMHV} ^{(1)\; LLA}&\simeq&  \frac{ i \pi }{1+w^*} \tilde{f}_1(w, w^*) +   \frac{ i \pi }{1+\frac{1}{w^*}}  \tilde{f}_1\left(\frac{1}{w}, \frac{1}{w^*} \right),
\eeqn
where
\beqn
\tilde{f}_1(w, w^*) =\log |1+w|^2.
\eeqn
The functions  $1/(1+w^*)$ and $1/(1+\frac{1}{w^*})$ are  related to $R$-invariants and are universal for all loops. Thus the problem of calculating the ratio function reduces to finding $\tilde{f}_{\ell}(w, w^*)$, where $\ell$ is the number of loops. By virtue of eq.~(\ref{PLLA}),  the function   $\tilde{f}_{\ell}(w, w^*)$ includes the MHV remainder function and it is useful  to  introduce a redefined  function $ f_{\ell}(w, w^*)$ defined by
\beqn
f_{\ell}(w, w^*) =\tilde{f}_{\ell}(w, w^*)-f^{MHV}_{\ell}(w, w^*),
\eeqn
where  $f^{MHV}_{\ell}(w, w^*)$ is the corresponding MHV contribution.
 The function $  f_{\ell}(w, w^*)$ can be found from  $f^{MHV}_{\ell}(w, w^*)$
using the property of the leading logarithm NMHV remainder functions in eq.~(\ref{diffeq})
\beqn \label{diffeqapp}
R_{NMHV}=-\int d w  \frac{w^*}{ w} \frac{\partial }{\partial  w^*} R_{MHV}
\eeqn
and demanding singlevaluedness and proper collinear behavior.
At two loops  we read out from eq.~(\ref{R26})
\beqn
f^{MHV}_{2}(w, w^*)=\log |1+w|^2 \log \left|1+\frac{1}{w}  \right|^2
\eeqn
and then using eq.~(\ref{diffeqapp}) obtain
\beqn\label{diffR2}
-\int d w  \frac{w^*}{ w} \frac{\partial }{\partial  w^*} f^{MHV}_{2}(w, w^*)
&& = \frac{1}{1+w^*} \left(  \log |w|^2 \log (1+w^*) -\frac{1}{2} \log^2 (1+w^*) -\text{Li}_2 (- w) \right) \nonumber \\
&& \hspace{-4cm }+
\frac{w^*}{1+w^*} \left( \frac{1}{2}\log^2 w  +\log w^* \log w -  \log w \log (1+w^*) + \text{Li}_2 (- w) \right)
+F (w^*), \;\;\;\;
\eeqn
where $F (w^*)$ is some  arbitrary function of only $w^*$.
We fix $F (w^*)$ by demanding singlevaluedness for $w$ being rotated by an arbitrary phase $\phi$ and $w^*$ rotated by $-\phi$. The best way to see how this determines  $F (w^*)$ is  to inspect the first term in eq.~(\ref{diffR2}), namely the function
\beqn
  \log |w|^2 \log (1+w^*) -\frac{1}{2} \log^2 (1+w^*) -\text{Li}_2 (- w).
\eeqn
Its symbol reads
\beqn\label{symb2raw}
&& w \otimes (1+w^*) +(1+w) \otimes w  +w^* \otimes (1+w^*)+(1+w^*) \otimes w
 \\
 &&
 +(1+w^*) \otimes w^* -(1+w^*) \otimes (1+w^*),   \nonumber
\eeqn
and the analytic continuation is done clipping the first entry~\cite{Dixon3loops}. In particular, for $w <1$
\beqn
w \otimes (1+w^*) \rightarrow w \otimes (1+w^*)+ i \phi  \;  (1+w^*)
\eeqn
and the last term cancels in eq.~(\ref{symb2raw}) against
\beqn
w^* \otimes (1+w^*) \rightarrow w^* \otimes (1+w^*)- i \phi  \;  (1+w^*).
\eeqn
For $w >1$ we have also
\beqn
(1+w) \otimes w \rightarrow (1+w) \otimes w+ i\phi  \;   w+\frac{(i \phi)^2}{2}
\eeqn
and  the last two terms  cancel against
\beqn
(1+w^*) \otimes w \rightarrow (1+w^*) \otimes w- i\phi  \;   w-\frac{(i \phi)^2}{2}.
\eeqn
This cancelation does not happen for $(1+w^*) \otimes w^*$ and $-(1+w^*) \otimes (1+w^*)$ in  eq.~(\ref{symb2raw}), and we can use the  freedom of choosing $F (w^*)$ to remove  those.
Thus we are left with
\beqn\label{symb2raw1}
  w \otimes (1+w^*) +(1+w) \otimes w  +w^* \otimes (1+w^*)+(1+w^*) \otimes w,
\eeqn
which matches
\beqn
  f_2 (w,w^*)=\log |w|^2 \log (1+w^*)  -\text{Li}_2 (- w)+\text{Li}_2 (- w^*)
\eeqn
up to a constant, which can be shown to be zero by demanding eq.~(\ref{symb2raw1}) to  be vanishing  as $|w| \to 0$ in the collinear limit.
Now we readily write the answer for the leading logarithm remainder function at two loops
\beqn
&& R^{(2)\; LLA}_{NMHV} \simeq \frac{i \pi }{2} \log \left(\frac{s_{23}}{s_0}\right) \left\{
 \frac{1}{1+w^*}   f_2 (w,w^*)
+\frac{1}{1+\frac{1}{w^*}}  f_2\left(\frac{1}{w}, \frac{1}{w^*} \right) \right\}.
\eeqn
We checked this result by a direct calculation using the Cauchy theorem.

We apply the same procedure to the three loop NMHV remainder function. Firstly, we read out from eq.~(\ref{R36})
\beqn\label{f3app}
&&f^{MHV}_3 (w,w^*)=
  \log |w|^2 \log^2 |1+w|^2 -\frac{2}{3} \log^3 |1+w|^2  -\frac{1}{4} \log^2 |w|^2 \log |1+w|^2 \nonumber
  \\
     &&
\hspace{3cm}
  +\frac{1}{2} \log |w|^2 \left( \text{Li}_2 \left(-w\right) +\text{Li}_2 \left(-w^*\right) \right)
 -\text{Li}_3 \left(-w\right)-\text{Li}_3 \left(-w^*\right)
\eeqn
and then calculate
\beqn\label{diff3}
-\int d w  \frac{w^*}{ w} \frac{\partial }{\partial  w^*} f^{MHV}_{3}(w, w^*)&& =
\frac{1}{1+w^*} \left\{
\frac{1}{2} \log |w|^2 \log (1+w^*) -\log (-w) \log |1+w|^2  \nonumber
\right.
\\
&&  \hspace{-5cm}
\left.
-\frac{1}{2} \log |w|^2  \; \text{Li}_2 (-w)-2 \log |1+w|^2 \text{Li}_2 (1+w)+\frac{1}{2} \log w  \;\text{Li}_2 (-w^*)+
2 \text{Li}_3 (1+w)+F (w^*)  \nonumber
\right\}
\\
&&  \hspace{-5cm} +\frac{w^*}{1+w^*} \left\{ \frac{}{} ... \right\},
\eeqn
where the last term is irrelevant for the present discussion because it can be found as the function which multiplies $1/(1+w^*)$.
Analyzing  the symbol of
\beqn
&& \frac{1}{2} \log |w|^2 \log (1+w^*) -\log (-w) \log |1+w|^2
-\frac{1}{2} \log |w|^2  \; \text{Li}_2 (-w)
\\
&& -2 \log |1+w|^2 \text{Li}_2 (1+w)+\frac{1}{2} \log w  \;\text{Li}_2 (-w^*)+
2 \text{Li}_3 (1+w)  \nonumber
\eeqn
we see that to ensure the singlevaluedness of the expression one should  add to it the following symbol
\beqn
-\frac{1}{2} w^* \otimes (1+w^*) \otimes w^* -2 w^* \otimes (1+w^*) \otimes (1+w^*)-(1+w^*) \otimes w^* \otimes w^*,
\eeqn
 which corresponds to
 \beqn
 \frac{1}{2} \log w^* \; \text{Li}_2 (-w^*) +2  \;  \text{Li}_3 (1+w^*).
 \eeqn
This determines $F (w^*)$ in eq.~(\ref{diff3}) up to a constant, which is fixed to  by demanding the entire  expression to vanish in the collinear limit $|w| \to 0$. Thus we have
\beqn
F (w^*)=\frac{1}{2} \log w^* \; \text{Li}_2 (-w^*) +2  \;  \text{Li}_3 (1+w^*) -4 \zeta_3
\eeqn
and  then
\beqn
f_3(w,w^*)=&& -\frac{1}{2} \log^2 |w|^2 \log(1+w^*)+\log (-w) \left( \log^2 (1+w^*)-\log^2 (1+w) \frac{}{}\right) \\
&& +2 \zeta_2 \log |1+w|^2 +\frac{1}{2} \log |w|^2 \left( \text{Li}_2 (-w)-\text{Li}_2 (-w^*)\frac{}{}\right) \nonumber
\\
&&
-2 \log |1+w|^2 \text{Li}_2 (-w) -2 \text{Li}_3 (1+w)-2 \text{Li}_3 (1+w^*)+ 4 \zeta_3.  \nonumber
\eeqn
for the NMHV remainder function at three loops in the leading logarithm approximation
\beqn
R^{(3)\; LLA}_{6;NMHV}  && \simeq \frac{i \pi }{4} \log^2 \delta \left(
\frac{1}{1+w^*} f_3(w,w^*)
+
\frac{1}{1+\frac{1}{w^*}} f_3\left(\frac{1}{w},\frac{1}{w^*}\right)
\right).
\eeqn

\end{document}